\documentstyle[aps,prd,floats,epsf]{revtex}
\tighten
\draft
\begin{document}

\twocolumn[\hsize\textwidth\columnwidth\hsize\csname
@twocolumnfalse\endcsname
\title{Black Hole Formation in Core-Collapse Supernovae and
Time-of-Flight Measurements of the Neutrino Masses}
\author{J.~F. Beacom\thanks{Electronic address:
        {\tt beacom@fnal.gov}}}
\address{Physics Department 161-33, California
Institute of Technology, Pasadena, California 91125}
\address{NASA/Fermilab Astrophysics Center, 
Fermi National Accelerator Laboratory, Batavia, Illinois 60510-0500}
\author{R.~N. Boyd\thanks{Electronic address:
        {\tt boyd@mps.ohio-state.edu}}}
\address{Departments of Astronomy and Physics, 
The Ohio State University, Columbus, Ohio 43210}
\author{A. Mezzacappa\thanks{Electronic address:
        {\tt mezz@nova.phy.ornl.gov}}}
\address{Physics Division,
Oak Ridge National Laboratory, Oak Ridge, Tennessee 37831}
\date{October 19, 2000; minor revisions January 8, 2001}
\maketitle

\begin{abstract}

In large stars that have exhausted their nuclear fuel, the stellar
core collapses to a hot and dense proto-neutron star
that cools by the radiation
of neutrinos and antineutrinos of all flavors.
Depending on its final mass, this may become either a neutron star or a
black hole.  Black hole formation may be triggered by mass accretion
or a change in the high-density equation of state.  We consider the
possibility that black hole formation happens when the flux of
neutrinos is still measurably high.  If this occurs, then the neutrino
signal from the supernova will be terminated abruptly (the transition
takes $\lesssim 0.5$ ms).  The properties and duration of the signal
before the cutoff are important measures of both the physics and
astrophysics of the cooling proto-neutron star.  For the event rates
expected in present and proposed detectors, the cutoff will generally
appear sharp, thus allowing model-independent time-of-flight
mass tests for the neutrinos after the cutoff.  If black hole
formation occurs relatively early, within a few ($\sim 1$) seconds
after core collapse, then the expected luminosities are of order
$L_{BH} = 10^{52}$ erg/s per flavor.  In this case, the neutrino mass
sensitivity can be extraordinary.  For a supernova at a distance $D =
10$ kpc, SuperKamiokande can detect a $\bar{\nu}_e$ mass down to 1.8
eV by comparing the arrival times of the high-energy and low-energy
neutrinos in $\bar{\nu}_e + p \rightarrow e^+ + n$.  This test will
also measure the cutoff time, and will thus allow a mass test of
$\nu_\mu$ and $\nu_\tau$ relative to $\bar{\nu}_e$.  Assuming that
$\nu_\mu$ and $\nu_\tau$ are nearly degenerate, as suggested by the
atmospheric neutrino results, masses down to about 6 eV can be probed
with a proposed lead detector of mass $M_D = 4$ kton (OMNIS).
Remarkably, the neutrino mass sensitivity scales as $(D/L_{BH}
M_D)^{1/2}$.  Therefore, {\it direct} sensitivity to all three
neutrino masses in the interesting few-eV range is realistically
possible; {\it there are no other known techniques that have this
capability}.

\end{abstract}

\vspace{0.25cm}]
\narrowtext


\section{Introduction}

In the past several years, the growing evidence for neutrino
oscillations has caused a great deal of excitement over the implied
nonzero neutrino masses.  Oscillation phenomena, however, are
sensitive only to differences of the squared neutrino masses, and thus
provide only a lower bound on the heavier mass.  Without further
input, the deduced masses can be increased, and the difference of
masses decreased, providing exactly the same difference of squared
masses and hence the same oscillation phenomena.

It is therefore of crucial importance to experimentally measure or
constrain the absolute scale of the neutrino masses.  Two indirect
techniques have been proposed.  First, the sum of the neutrino masses
can be constrained by cosmological arguments.
The requirement of not overclosing the universe gives an upper bound
of about 100 eV~\cite{cosmo1}.  This bound may be improved by 
considering the effects of neutrino masses on the cosmic microwave 
background and the clustering of galaxies; the claimed (future, in some 
cases) sensitivity is about 1 -- 10 eV~\cite{cosmo2}.  These arguments 
require that the other cosmological parameters are independently known 
and may not apply if the neutrinos decay.\footnote{Furthermore, it has 
recently been shown that in scenarios with a low (MeV-scale) reheating 
temperature, the neutrinos may decouple without reaching equilibrium, 
leading to a substantially lower density than in the usual scenario; 
this may weaken the cosmological neutrino mass bounds by a factor of 
10 or more~\cite{lowreheat}.}
Second~\cite{Barger}, if all of the neutrino masses are connected
by small measured mass-squared differences, then each mass is
constrained by the limit on the electron neutrino mass from tritium
beta decay, now about 3 eV~\cite{tritium} (the direct laboratory
limits on the mu and tau neutrino masses are 170 keV~\cite{Assamagan}
and 18 MeV~\cite{Barate}, respectively).  If neutrinoless double beta
decay were discovered (i.e., neutrinos were confirmed to have a
Majorana character), then this could anchor the masses at an even
lower value~\cite{2betamass}; the present limit on the combination 
of masses measured in double beta decay is about 0.2 eV~\cite{Baudis}.
Strictly speaking, to use the arguments of Ref.~\cite{Barger}, each
oscillation signal must first be decisively confirmed, including
precise measurement of the mixing parameters and identification of the
oscillated flavors.  Until then, we must allow for the possibility
that there are more relevant flavors than there are measured
mass-squared differences.  For example, if the solar neutrino problem
is solved by $\nu_e \rightarrow \nu_s$ oscillations, and if the LSND
signal is ruled out, then the atmospheric neutrino problem can be
solved by $\nu_\mu \rightarrow \nu_\tau$ oscillations with a small
mass difference and large masses, say 10 or 100 eV, as long as $\delta
m^2 \simeq 10^{-3}$ eV$^2$~\cite{SKatm}.

Thus, while the indirect constraints on the neutrino masses are
valuable, it would be much more satisfying to have a direct
experimental measurement.  Presently, the best possibility for direct
measurement of the mu and tau neutrino masses is by time-of-flight
differences using neutrinos from a Galactic core-collapse supernova.
\footnote{As noted by Shrock~\cite{Shrock}, if neutrinos are mixed,
then beta decay spectra consist of incoherent contributions from each
mass eigenstate, where the endpoints depend on the masses, and the
weighting on the mixing angles.  The presence of kinks in the spectrum
would thus allow direct measurement of $m_2$ and $m_3$ and their
mixing angles.  In order to experimentally separate such kinks from
an endpoint turnover due to $m_1$, the mass differences and the
mixing angles must be large enough.  For light neutrinos, the 
$\bar{\nu}_e$ disappearance experiments presently provide more 
restrictive limits on these parameters.}
At lowest order, a neutrino with mass $m$ (in eV) and energy $E$ (in
MeV) will experience an energy-dependent delay (in s) relative to a
massless neutrino in traveling over a distance D (in 10 kpc):
\begin{equation}
\Delta t(E) = 0.515 \left(\frac{m}{E}\right)^2 D\,.
\label{eq:delay}
\end{equation}
The distance is scaled by the approximate distance to the Galactic
center, though a supernova may be detected from anywhere in the Galaxy
and its immediate companions (e.g., the Magellanic Clouds).
SuperKamiokande (SK) and the Sudbury Neutrino Observatory (SNO) would
have good sensitivity to a Galactic supernova, collecting of order
$10^4$ and $10^3$ events, respectively (see
Refs.~\cite{SKpaper,SNOpaper,JHUproc} and references therein).  Unless
the decreasing neutrino luminosity is interrupted by black hole
formation, it should be possible to measure it to very late times
(some tens of seconds); either outcome would be an important probe of
the nuclear equation of state~\cite{Janka,Pons}.

The primary interest for mass tests is to measure the mu and tau
neutrino masses relative to the nearly massless electron neutrino.  A
neutrino mass test~\cite{SKpaper,SNOpaper} based on the average event
arrival times $\langle t \rangle$ can measure a mu or tau neutrino
mass as small as 45 eV in SK and 30 eV in SNO.  If the mu and tau
neutrinos are maximally mixed with nearly degenerate masses, then the
sensitivity on either mass eigenstate is better by a factor of about
$\sqrt{2}$, i.e., about 30 eV in SK and 20 eV in
SNO~\cite{SKpaper,SNOpaper}.  This test is independent of
supernova neutrino emission models, though it does assume that the
luminosities of the different flavors have similar {\it shapes} as a
function of time, as expected on general grounds and also seen in the
supernova models~\cite{SNmodels}.  In the absence of a model, a
Kolmogorov-Smirnov test can be used to compare the event rates for
different flavors of neutrinos; it can be shown that this reduces to
the $\langle t \rangle$ test~\cite{triang}.  Other tests proposed in
the literature are explicitly model-dependent, and the models have
large uncertainties.

While the $\langle t \rangle$ test could improve the limit on the tau
neutrino mass by almost six orders of magnitude, it seems very
difficult to reach the eV range suggested by the cosmological and
tritium arguments above.  It can be shown~\cite{SNOpaper} that the
mass sensitivity generically scales with the detector mass $M_D$ as
$1/M_D^{1/4}$; therefore, another order of magnitude in sensitivity in
neutrino mass would require detectors $10^4$ times larger, which seems
impossible.  It can also be shown~\cite{SNOpaper} that the sensitivity
is {\it independent} of the distance to the supernova in the case
where the deduced neutrino mass is compatible with zero and only an
upper limit is placed.

In this paper, a comprehensive study that follows our recent {\it
Letter}~\cite{BHprl}, we consider the case that the proto-neutron star
forms a black hole, instead of gradually cooling as a stable neutron
star.  If that happens early enough, then the neutrino signals will be
abruptly terminated as the neutrinospheres are enveloped by the event
horizon of the black hole.  In Section II, we discuss the conditions
required for this to happen and to be observable, as well as the
expected details of the neutrino signal.  In Section III, we derive
the mass effects on the detected neutrino event rate in the general
case.  In Section IV, we show how to measure the black hole cutoff
time in SuperKamiokande, with or without the complicating effects of a
possible electron neutrino mass.  In Section V, we show how to make a
time-of-flight mass measurement of the mu and tau neutrino masses
relative to the cutoff time measured in SK.  Finally, in Section VI,
we discuss some remaining issues and conclude.


\section{Black Hole Formation and the Supernova Neutrino Signal}

Before discussing how to measure the neutrino masses, we first examine
how likely it is that black hole formation will truncate the neutrino
flux from a Galactic supernova.  Three questions naturally arise:

(1) Is the Galactic supernova rate reasonably high?

(2) Are black holes formed reasonably often in core-collapse supernovae?

(3) Can black hole formation occur when the neutrino fluxes are still
high?

An examination of the evidence reveals that, while the uncertainties
are large, there is a good chance of satisfying all three
requirements.  If so, this could have a profound impact on our ability
to {\it directly} measure all three neutrino masses.  Before showing
how that could be done, we address these requirements.


\subsection{Galactic Supernova Rate}

From studies of other galaxies, we know that about 80--90\% of
supernovae are of the core-collapse type (types II, Ib, Ic), which
produce a substantial flux of neutrinos~\cite{SNrate,SNhist}.  In the
following we treat the overall supernova rate without regard to
correction for the smaller rate of Ia supernovae.

A rough estimate of the Galactic supernova rate can be made using the
historical records.  Over the past 1000 years, 7 Galactic supernovae
are known either from historical records or their
remnants~\cite{SNhist,newSN}.  Probably some others in the southern
sky were missed because they were not visible to or not recorded by
the astronomers of the time.  For example, the recently-discovered
supernova remnant reported in Ref.~\cite{newSN} is apparently
extremely close (0.2 kpc) and only about 700 years old, but is not
found in the historical record.  It is therefore not unreasonable to
estimate that nearby supernovae occur at a rate of about 1/century.
Due to obscuration by dust, naked-eye supernovae are not visible
beyond several kpc (the farthest of these 7 was at 4.2 kpc);
therefore, one must correct for the small fraction of the Galaxy
surveyed.  The Bahcall-Soneira Galactic model~\cite{Bahcall1,Bahcall2}
includes somewhat less than 10\% of the stars within about 4 kpc of
Earth; therefore, we estimate the total Galactic supernova rate to be
about 10/century (see also Refs.~\cite{Strom,Hatano}).

This estimate of 10/century agrees with the rate given by Bahcall and
Piran~\cite{Bahcall2}, who make a direct integration over the stellar
initial mass function, corresponding stellar lifetimes, and spatial
distribution of stars; their calculation is {\it not} normalized to
the historical rate.  It also agrees with the nucleosynthesis
arguments of Arnett, Schramm, and Truran~\cite{Arnett}.

On the other hand, more conservative estimates suggest that the rate
is lower: $(3 \pm 1)$/century~\cite{SNrate,SNhist}.  It is not clear
how to reconcile this with the above estimates of 10/century.  The
estimate based on the historical rate and the independent
Bahcall-Piran calculation agree, and the only element they have in
common is the fraction of stars nearby.  Thus, the most likely fault
with these calculations, if any, is that they assume that the stars
that explode as supernovae are distributed in the same way as other
stars.  In fact, Refs.~\cite{SNreduce1,SNreduce2} argue against this
assumption, and claim that the nearby supernova rate is anomalously
high due to our occupying a privileged position in the Galaxy.

With coverage over most of the Galaxy over most of the past 20 years,
no neutrino detectors have reported a Galactic
supernova~\cite{SNsearch} (note that SN1987A is excluded because it
occurred in the Large Magellanic Cloud).  Taken at face value, this
would exclude a Galactic supernova rate of 10/century at about the
85\% CL.  However, an analysis combining all of the experiments has
not been done, and is needed.  A number of these experiments did not
have full coverage of the Galaxy and/or had significant ($\simeq
50\%$) downtime, and taking this into account will yield a weaker
constraint.

LIGO~\cite{LIGO} and other novel techniques~\cite{Dupraz,Diehl,Haxton}
may also be able to shed some new light on the supernova rate.

The combined evidence thus suggests a Galactic supernova rate of at
least 3/century.


\subsection{Relative Frequency of Black Hole Formation}

From a theoretical point of view, the relative frequency of black hole
and neutron star formation (the BH/NS ratio) depends on the equation
of state of nuclear matter~\cite{nucmatter} and the supernova
mechanism~\cite{SNmodels}; further work on each is greatly needed.
Ideally, appropriate direct observational constraints on neutron-star
properties could be decisive for discriminating between different
equations of state~\cite{nucmatter,NSobs}.

As is well-known, SN1987A in the Large Magellanic Cloud ($D \simeq 50$
kpc) was clearly observed by the Kamiokande II and IMB detectors, with
12 and 8 events, respectively~\cite{KamII,IMB}.  The observed duration
of SN1987A was about 10 s, consistent with a supernova that formed a
neutron star.  No neutron star has been seen yet in the remnant, but
this may not mean that one is not present~\cite{Zampieri}.  Thus if a
black hole formed, it evidently happened after the neutrino flux died
out~\cite{BH87A}.  The progenitor mass plays an important role in
deciding the ultimate mass and hence fate of the core.  Thus, even
though SN1987A (progenitor mass $\sim 18 M_\odot$) did not form a
black hole in the first 10 s after collapse, other supernovae will be
different.

Core-collapse supernovae occur only for stars massive enough to burn
their cores up to iron; this minimum mass is estimated to be about $8
M_\odot$.  It is also generally believed that stars above some mass,
perhaps $20 M_\odot$, will always produce black holes instead of
neutron stars.  Bahcall and Piran~\cite{Bahcall2} estimate that
supernovae from progenitors above $20 M_\odot$ number about 1/2 of
those below $20 M_\odot$.  Ratnatunga and van den
Bergh~\cite{Ratnatunga}, with a supernova rate several times smaller,
estimate about 1/4 for this ratio.  Fryer~\cite{Fryer} estimates a
BH/NS ratio somewhere between a few percent and 1/4, depending on the
cutoff progenitor mass; both are strongly affected by the
uncertainties in the inputs to his supernova code.

For an assumed stellar initial mass function, predictions of the
remnant mass distribution have been made by Timmes, Woosley, and
Weaver~\cite{Timmes}, who find a bimodal distribution with peaks at
about $1.3 M_\odot$ and $1.8 M_\odot$.  (In some other models, this
bimodal distribution is not seen; see, e.g., Fig.~3 of
Ref.~\cite{Umeda}).  The bimodal nature in this model is due to
progenitor masses below and above $19 M_\odot$, which either burn
carbon convectively or radiatively, respectively~\cite{Timmes}.  If
the maximum neutron star mass is the conventional $2.2
M_\odot$~\cite{Akmal}, then the BH/NS ratio $\simeq 0$~\cite{Timmes}.
However, Brown and Bethe~\cite{Brown94,Bethe95} argue that the maximum
neutron star mass is about $1.5 M_\odot$, on the basis of both an
assumed softer equation of state and a number of observational
constraints.  In the Brown and Bethe model, progenitors above about
$18 M_\odot$ will form black holes, and they independently deduce a
BH/NS ratio $\sim 1$.  For a maximum neutron star mass of $1.5
M_\odot$, the Timmes, Woosley, and Weaver remnant distribution
indicates a BH/NS ratio $\simeq 3$ (the upper peak is larger than the
lower peak)~\cite{Timmes}.

Recent results by Ergma and van den Heuvel~\cite{Ergma} indicate that
the vast majority of progenitors above $20 - 25 M_\odot$ produce black
holes (this therefore supports a much higher BH/NS ratio than the
earlier paper of van den Heuvel and Habets~\cite{vandenHeuvel} that
suggested a BH/NS ratio $\simeq 1/100$).  This is corroborated by
Ref.~\cite{Zwart}, which suggests that the progenitor mass cutoff may
be even lower.  These results thus suggest a high BH/NS ratio.

It may eventually be possible to address the absolute BH/NS ratio
observationally via the BH/NS ratio (perhaps as large as 10 in a
preliminary study~\cite{Grindlay}) deduced from low-mass x-ray
binaries, though this also depends on the details of the binary
evolution.

Qian, Vogel, and Wasserburg~\cite{Qian} assumed that the $r$-process
production of heavy nuclei occurs in core-collapse supernovae and
considered the effects of black hole versus neutron star formation on
the yields.  They found that the observed $r$-process distribution may
be best explained with a very high BH/NS ratio $\sim 10$.  Their
results require that black hole formation happens early, when
the neutrino fluxes are relatively high, which will terminate part of
the $r$-process production.  While their BH/NS ratio is very large,
their hypothesis is supported by recent measurements~\cite{Sneden}.
Further measurements of $r$-process yields in ultra-metal-poor stars
would be very valuable.

The accumulated evidence thus supports a relatively high BH/NS ratio,
so that the next Galactic supernova would be likely to form a black
hole.


\subsection{Scenarios for Black Hole Formation}

One scenario for black hole formation in core-collapse supernovae
occurs if the proto-neutron star mass exceeds the maximum neutron star
mass.  For ordinary neutron-rich nuclear matter, this maximum mass is
thought to be about $2.2 M_\odot$~\cite{Akmal}, though there may be
significant uncertainties.  This may occur in the initial collapse, or
after some delay, due to accretion of further mass.  The neutrino
signal expected in a scenario of this type has been studied by
Burrows~\cite{Burrows88} and Mezzacappa and Bruenn~\cite{Mezzacappa}
(see also a very early paper by Wilson~\cite{Wilson}).
In these models, neutrino emission was followed until abruptly
terminated by black hole formation (the results do not continue
through the short but nonzero black hole formation time).  Before the
cutoffs at 1 -- 2 s, the luminosities were fairly constant at more
than $10^{52}$ erg/s per flavor.

A second scenario for black hole formation is based on a softening of
the equation of state in the proto-neutron star as the neutrinos are
emitted and a phase transition to a more exotic state of matter
occurs, containing perhaps strange mesons or baryons, charged-pion
condensates, or free quarks.  The maximum neutron star mass for such
exotic nuclear matter is generally
lower~\cite{Brown94,Bethe95,NSmass}, perhaps about $1.5 M_\odot$.
Thus an initially stable proto-neutron star may form a black hole
after the phase transition.  The details of the neutrino signal
accompanying black hole formation in such scenarios have been studied
by Baumgarte et al.~\cite{Baumgarte96b}; see also earlier
work~\cite{Keil,Glendenning,Baumgarte96a}.  A detailed study in full
general relativity was made of the neutrino emission just before and
during the formation of the black hole.  A singularity-avoiding
code~\cite{Baumgarte95} was used that tracked the emission in the
frame of a distant observer (i.e., the result is the redshifted,
time-dilated luminosity that would be seen in a neutrino detector).
Before the cutoff at about 10 s, the luminosities were fairly constant
at about $10^{51}$ erg/s per flavor.

Finally, we discuss two scenarios in which a neutron star can become a
black hole long after the neutrino flux has died away.  As such, these
scenarios are not of direct interest to us.

In a successful supernova explosion, the outgoing shock will pass
through the stellar envelope within a few hours after core collapse.
If a reverse shock forms, it may dump matter onto the
neutron star and cause it to exceed the maximum mass, hence
causing black hole formation.  These fallback scenarios are discussed
in Refs.~\cite{Zampieri,fallback}.

Gradual accretion onto old neutron stars until the maximum mass is
exceeded is also possible, and a concrete scenario is discussed by
Gourgoulhon and Haensel~\cite{Gourgoulhon}.  Given specific
assumptions about the equation of state of nuclear matter, they find
that in the last stages of accretion that the matter will become less
neutron-rich, and will emit a burst of $\bar{\nu}_e$ neutrinos with
$\langle E \rangle \simeq 3$ MeV.  This lasts $\simeq 0.5$ ms until
truncated by black hole formation.  At 10 kpc, we estimate that this
would cause $\sim 3$ events above the SK threshold.  In fact, since
their model does not include neutrino opacities, the neutrino energy
and the luminosity before the cutoff will both be lower.  Thus, unless
the neutron star is very close, this would be undetectable (see also
Ref.~\cite{Fulgione} for a study of the sensitivity of LVD).

Thus there are some concrete
models~\cite{Burrows88,Mezzacappa,Baumgarte96b} in which black hole
formation occurs early enough to cut off the neutrino fluxes when
they are still measurably high, though the uncertainties are large and
depend on the details of the supernova models.


\subsection{Details of the Neutrino Signal}

In the general case, the observables for each neutrino flavor are the
luminosity $L(t)$ and temperature $T(t)$ up to and during the time of
black hole formation.  The duration of the cutoff must be very short,
of order the light crossing time $2R/c \simeq 0.1$ ms.  In the most
detailed numerical treatment available~\cite{Baumgarte96b}, the
duration of the cutoff is about 0.5 ms.  We assume that
this will be typical for {\it any} mechanism of black hole
formation.  For black hole formation at very early times, the initial
proto-neutron star would be larger than assumed in
Ref.~\cite{Baumgarte96b}, and one might argue that this would lengthen
the duration of the cutoff.  However, it should be noted that what
defines the cutoff is the increasing gravitational redshift, and this
does not become large until the proto-neutron star is already very
compact.  For emission from the proto-neutron star, the neutrino
gravitational redshifts are moderate; $z \simeq G M/R c^2 \sim 0.1$.
The redshifts only become severe during the short cutoff at $t_{BH}$,
when $z \rightarrow \infty$ (using the full expression for $z$).  In
any case, further modeling of the neutrino signal up to and during
black hole formation is needed, especially for black hole formation at
earlier times.  It will be shown below that the statistical error in
defining the position of the cutoff is larger than 0.5 ms; therefore,
all of the neutrino flavors can be considered to be cut off sharply
and simultaneously at a time $t_{BH}$.  These approximations can be
made because the expected numbers of events during the cutoff are less
than 1.

In Ref.~\cite{Baumgarte96b}, some interesting details of the signal
during the $\simeq 0.5$ ms cutoff are pointed out.  The very last
neutrinos to be seen will not come from radial paths, but rather from
unstable circular orbits.  It should be noted that the calculation of
Ref.~\cite{Baumgarte96b} only treats neutrinos on radial paths.  The
final decay of the luminosity due to the neutrinos on unstable
circular orbits is expected to be exponential, with a time constant
proportional to the black hole mass~\cite{spiral}.  Since this time
constant is very small, $\tau = 3\sqrt{3} G M_{BH}/c^3 \simeq 0.04$
ms, the number of such events (proportional to the disregarded
luminosity multiplied by this duration) will be negligible for the
cases considered in this paper.  Normally, electron neutrinos are
emitted from the largest radius and with the lowest temperature.  At
the end of the neutrino signal during black hole formation, the
electron neutrinos will be cut off last and will briefly have a higher
temperature than the other flavors (due to less gravitational
redshift).  Unfortunately, all of these details of the transition are
not observable with the present and proposed detectors, due to the
limited statistics.  For a very close supernova, the situation might
be different; this will be discussed below.

The abrupt and simultaneous termination of all flavors of neutrinos
allows a very simple mass test.  Since the electron neutrino is nearly
massless, the termination of the $\bar{\nu}_e$ event rate in SK will
signal the black hole formation time $t_{BH}$ (the effects of a
possible electron neutrino mass will be discussed below).  Then, any
events observed after $t_{BH}$ could only have come from
neutral-current detection of time-of-flight delayed, massive
$\nu_\mu$, $\nu_\tau$, $\bar{\nu}_\mu$, and $\bar{\nu}_\tau$.  We have
assumed that the detector background is negligible, in the sense that
the expected number of background events over a typical delay time is
$\ll 1$.

Before $t_{BH}$, one would like to measure $L(t)$ and $T(t)$ for all
of the neutrino flavors.  This is straightforward for $\nu_e$ and
$\bar{\nu}_e$, since the detected outgoing lepton carries nearly the
full neutrino energy in reactions with nuclear targets.  Since
$\nu_\mu$, $\nu_\tau$, $\bar{\nu}_\mu$, and $\bar{\nu}_\tau$ only have
enough energy to undergo neutral-current interactions, they are
indistinguishable.  However, for the same reason, they are also
expected to be produced with the same luminosity and temperature.  It
is not generally possible to measure the temperature for these species
directly, and it must be inferred by the yields on different targets
(cross sections with different energy dependence sample the spectrum
differently; see Fig.~3 of Ref.~\cite{JHUproc}).  The measurements of
$L(t)$ and $T(t)$ for the various flavors before $t_{BH}$, as well as
the value of $t_{BH}$ itself, are important probes of the supernova
mechanism and the equation of state~\cite{Janka,Pons}.  They will also
be important for measuring the quantities needed for the mass
measurement, in order to reduce the model dependence.

In the bulk of this paper, we concentrate in the analysis on mu and
tau neutrino masses near the limit of detectability.  The mass effects
will then not appreciably affect the time dependence of the event rate
except at the sharp cutoff at $t_{BH}$.  In fact, it will be shown
that only the luminosity and temperature at $t_{BH}$ itself are
relevant.  In the models~\cite{Burrows88,Mezzacappa,Baumgarte96b}
considered, the neutrino luminosities and temperatures before $t_{BH}$
are roughly constant over the time scales of relevant mass delays.
Thus it is adequate (and much more convenient analytically) to
consider that the luminosities and temperatures of all flavors are
constant for some period before the cutoff, and that they have
simultaneous step-function cutoffs at $t_{BH}$.  These assumptions
will be relaxed below.  

We assume the following temperatures: $T = 3.5$ MeV for $\nu_e$, $T =
5$ MeV for $\bar{\nu}_e$, and $T = 8$ MeV for $\nu_\mu$, $\nu_\tau$,
$\bar{\nu}_\mu$, and $\bar{\nu}_\tau$~\cite{SNmodels}.  This hierarchy
is a consequence of the different opacities in the proto-neutron star,
and the decreasing temperature with increasing radius.  The
temperatures in Ref.~\cite{Baumgarte96b} were somewhat higher than
these conventional values, but the authors explain that this is probably due
to a numerical approximation in the transport code.

To illustrate our results quantitatively, we present results for two
concrete cases.  In the first, called {\bf ``Early,''} black hole
formation is assumed to occur a few ($\sim 1$) seconds after core
collapse, when the neutrino luminosities are of order $10^{52}$ erg/s
per flavor.  This case is nominally associated with black hole
formation by accretion onto the proto-neutron
star~\cite{Burrows88,Mezzacappa}.  In the second, called {\bf
``Late,''} black hole formation is assumed to occur within several
($\simeq 10$) seconds after core collapse, when the neutrino
luminosities are of order $10^{51}$ erg/s per flavor.
\footnote{Recent work of Pons et al.~\cite{Pons} suggests that black hole
formation would occur after a few tens of seconds;
however, their final luminosities are comparable to what we assume.}
This case is nominally associated with black hole formation by a
softening of the high-density equation of state in the proto-neutron
star~\cite{Baumgarte96b}.
Direct extraction of the $\bar{\nu}_e$ luminosity from the SN1987A
data roughly supports the luminosity-time correspondences given here.
It should be remembered that these are just examples---it will be
shown that all of the necessary quantities can be {\it measured} in a
realistic situation.


\section{Neutrino Mass Effects}


\subsection{Detected Event Rate}

For a constant, normalized, thermal spectrum $f(E)$, but a general
luminosity $L(t)$, the event rate for neutrinos with nonzero mass is
\begin{equation}
\frac{dN}{dt} = \frac{N_T}{4\pi D^2} \frac{1}{\langle E \rangle}
\int_0^\infty dE\, f(E) \sigma(E) L(t - \Delta t(E))\,,
\label{eq:rate}
\end{equation}
where $N_T$ is the number of targets in the detector, $D$ the
supernova distance, and $\langle E \rangle$ the average energy (for a
Fermi-Dirac spectrum $\langle E \rangle = 3.15 T$).  Generalization to
a time-dependent spectrum or a shape more general than Fermi-Dirac
would be straightforward.  The argument of $L(t)$ is shifted to
account for the possible energy-dependent delay of a massive neutrino.

As discussed above, we assume that the luminosity and temperature are
constant before $t_{BH}$ (for at least much longer than the typical
delay time), and then vanish abruptly.  That is, $L(t) =
L_{BH}\,\theta(t_{BH} - t)$, where $L_{BH}$ is the luminosity at the
cutoff.  In Eq.~(\ref{eq:rate}), we need to evaluate this with the
delayed argument, i.e., $L(t - \Delta t(E)) = L_{BH}\,\theta(t_{BH} -
t + \Delta t(E))$.  For $t < t_{BH}$, the step function is satisfied
for all energies, and the event rate is
\begin{equation}
\frac{dN}{dt}(t) = C
\left[\frac{L_{BH}}{10^{51}{\rm erg/s}}\right]
\int_0^\infty dE\, f(E)
\left[\frac{\sigma(E)}{10^{-42}{\rm cm^2}}\right]\,.
\label{eq:ratebefore}
\end{equation}
The integral is the thermally-averaged cross section
\begin{equation}
\sigma_{eff} =
\int_0^\infty dE\, f(E) \sigma(E)\,.
\label{eq:sigmaeff}
\end{equation}
This constant event rate is the same for massless neutrinos; as long
as the delays are much less than the total duration of the supernova
signal and the luminosity is constant, then at a given time the number
lost by delays to later times is compensated by the number gained by
delays from earlier times.  This is not true at the start of the
neutrino signal, but the rise is much less sharp than the black hole
cutoff, is model-dependent, and is not considered further.  For $t >
t_{BH}$, there is an upper limit on the neutrino energy, which must be
small enough for the neutrino to be delayed that long after $t_{BH}$.
Then
\begin{equation}
\frac{dN}{dt}(t) = C
\left[\frac{L_{BH}}{10^{51}{\rm erg/s}}\right]
\int_0^{E_{max}} \! dE\, f(E)
\left[\frac{\sigma(E)}{10^{-42}{\rm cm^2}}\right].
\label{eq:rateafter}
\end{equation}
The upper limit $E_{max}$ is simply the energy that makes the argument
of the step function $\theta(t_{BH} - t + \Delta t(E))$ vanish; using
Eq.~(\ref{eq:delay}), this is
\begin{equation}
E_{max} = m \sqrt{\frac{0.515 D}{t - t_{BH}}}\,,
\label{eq:emax}
\end{equation}
where the units are as in Eq.~(\ref{eq:delay}).  Note that the
neutrino mass and time dependence appear only through the limit of
integration.  If the neutrino energy can be measured, as in some
charged-current reactions, then the event rates for separate ranges of
neutrino energy can easily be obtained.  For an H$_2$O detector, the
constant $C$ is
\begin{equation}
C_{\rm H2O} = (1.74/{\rm s})
\left[\frac{M_D}{1{\rm\ kton}}\right]
\left[\frac{10{\rm\ kpc}}{D}\right]^2
\left[\frac{1{\rm\ MeV}}{\langle E \rangle}\right]\,.
\label{eq:CH2O}
\end{equation}
The constant for a $^{208}$Pb detector can be obtained by scaling by
the relative number of targets/kton, i.e., 18/208; therefore
\begin{equation}
C_{208{\rm Pb}} = (0.151/{\rm s})
\left[\frac{M_D}{1{\rm\ kton}}\right]
\left[\frac{10{\rm\ kpc}}{D}\right]^2
\left[\frac{1{\rm\ MeV}}{\langle E \rangle}\right]\,.
\label{eq:208Pb}
\end{equation}


\subsection{Number Delayed Past $t_{BH}$}

The expected number of delayed counts $N_{del}$ after $t_{BH}$ can be
determined analytically by integration of Eq.~(\ref{eq:rateafter}),
which will be useful when $t_{BH}$ can be measured independently.  This
is simply
\begin{eqnarray}
N_{del}
& = & 
\int_{t_{BH}}^\infty dt \; \frac{dN}{dt}(t) \\ \nonumber
& = & 
C \left[\frac{L_{BH}}{10^{51}{\rm erg/s}}\right]
\int_{t_{BH}}^\infty dt \int_0^\infty dE\, \\ \nonumber
& & 
\times f(E)
\left[\frac{\sigma(E)}{10^{-42}{\rm cm^2}}\right]
\theta(t_{BH} - t + \Delta t(E)) \\ \nonumber
& = &
C \left[\frac{L_{BH}}{10^{51}{\rm erg/s}}\right]
\int_0^\infty \! dE\, f(E)
\left[\frac{\sigma(E)}{10^{-42}{\rm cm^2}}\right]
\Delta t(E).
\end{eqnarray}
Note that the upper limit on energy in Eq.~(\ref{eq:rateafter}) was
written using the step function $\theta(t_{BH} - t + \Delta t(E))$;
this step function then disappeared in the integration over $t$.  Now
define
\begin{equation}
\langle \Delta t(E) \rangle_{f\sigma} =
\frac{\int_0^\infty dE\, f(E) \sigma(E)\Delta t(E)}
{\int_0^\infty dE\, f(E) \sigma(E)}\,,
\label{eq:typdel1}
\end{equation}
where the $f \sigma$ subscript emphasizes that the weighting is over
$f(E) \sigma(E)$, and not $f(E)$ alone (as for $\langle E \rangle$).
Then
\begin{eqnarray}
N_{del} & = & 
\langle \Delta t(E) \rangle_{f\sigma} \nonumber \\
& \times &
C \left[\frac{L_{BH}}{10^{51}{\rm erg/s}}\right]
\int_0^\infty dE\, f(E) 
\left[\frac{\sigma(E)}{10^{-42}{\rm cm^2}}\right]
\end{eqnarray}
Recognizing the event rate before (or at) $t_{BH}$ from
Eq.~\ref{eq:ratebefore}, this becomes
\begin{equation}
N_{del} =
\frac{dN}{dt}(t_{BH}) \times \langle \Delta t(E) \rangle_{f\sigma}\,.
\label{eq:ndel1}
\end{equation}
By use of Eq.~(\ref{eq:delay}), we see that Eq.~(\ref{eq:typdel1})
simply defines the average value of $1/E^2$.  By the mean-value
theorem for integrals, this can be written as $1/E_c^2$, where $E_c$
is a constant to be determined.  The weighted delay can then be
expressed as
\begin{equation}
\langle \Delta t(E) \rangle_{f\sigma}
= 0.515 \left(\frac{m}{E_c}\right)^2 D\,.
\label{eq:typdel2}
\end{equation}
The physical significance of the ``central'' energy $E_c$ is that it
is (to an excellent approximation) simply the Gamow peak of the
falling thermal spectrum and the rising cross section.  It can also be
determined by numerical evaluation of Eq.~(\ref{eq:typdel1}).  Thus we
arrive at the very simple and important result:
\begin{equation}
N_{del} =
\frac{dN}{dt}(t_{BH}) \times 0.515 \left(\frac{m}{E_c}\right)^2 D\,,
\label{eq:ndel1.5}
\end{equation}
where the event rate is in s$^{-1}$, and the other units are as in
Eq.~(\ref{eq:delay}).  This formula would obviously be true if only a
single energy contributed and the sharp cutoff in the event rate (see
Fig.~(\ref{fig:rateE}) and Fig.~(\ref{fig:rateL}) below) were simply
rigidly translated by the delay.  But it is remarkable and very
convenient that it is still true even when there is a spectrum of
energies and the event rate develops a decaying tail past the cutoff.
As derived, this is an exact result.

In the derivation of these results we assumed that the luminosity and
temperature (and hence also the event rate) were nearly constant
before $t_{BH}$, as suggested by the results of
Ref.~\cite{Burrows88,Mezzacappa,Baumgarte96b}.  For an arbitrary event
rate, a fit can always be made to the event rate before
$t_{BH}$, and $dN/dt$ at $t_{BH}$ extracted and used in the formula
for $N_{del}$.  (Below, we also discuss how $T$ and hence $E_c$
can be extracted from the data).
To integrate $N_{del}$ as above, it is only necessary
that the event rate be approximately constant over the scale of the
small possible mass delays, which is a very mild assumption.

Once the other quantities can be measured, then the neutrino mass $m$
is given by Eq.~(\ref{eq:ndel1.5}).  We show how this can
be done below.


\section{Measurement of the Electron Neutrino Mass}


\subsection{CC Event Rate and Measurement of $t_{BH}$}

We first consider how well $t_{BH}$ could be measured if we knew that
$m_{\nu_e} \simeq 0$.  The dominant event rate in SK is from
$\bar{\nu}_e + p \rightarrow e^+ + n$.  The cross
section~\cite{invbeta} as a function of the neutrino energy $E$,
including the recoil, weak magnetism, and radiative corrections, is
well-approximated at typical supernova neutrino energies (where we can
disregard the electron mass) by
\begin{equation}
\sigma(E) =
0.0952 \, (E - 1.3)^2 \, (1 - 7 E/M)\,,
\label{eq:invbeta}
\end{equation}
for neutrino energies $E > 1.8$ MeV.  In this formula, $M$ is the
nucleon mass in MeV, and the cross section is in $10^{-42} {\rm\
cm}^2$.  For a temperature $T = 5$ MeV, the thermally-averaged cross
section (for the {\it sum} of the two protons in H$_2$O) is
$44. \times 10^{-42} {\rm\ cm}^2$.  (This is slightly smaller than the
result used in Refs.~\cite{SKpaper,oxygen}, due to an improved
treatment of the corrections~\cite{invbeta}).  Thus, for a supernova
at 10 kpc as seen in SK (32 kton), the event rate due to $\bar{\nu}_e
+ p \rightarrow e^+ + n$ can be easily calculated.  Using
Eq.~(\ref{eq:ratebefore}), the rate just before the cutoff at $t_{BH}$
is $\simeq 1500 {\rm\ s}^{-1}$ in the Early case and $\simeq 150 {\rm\
s}^{-1}$ in the Late case.  After $t_{BH}$, the rates are zero.  We
have disregarded the 0.5 ms duration~\cite{Baumgarte96b} of the cutoff,
which should contain about 0.4 events in the Early case and about 0.04
events in the Late case.  Since these are fewer than 1, the cutoff can
be considered to be sharp.

How is the cutoff time measured, and what is its error?  Suppose we
have an event rate $R(t)$ measured before the unknown cutoff time
$t_{BH}$.  The time of the last event $t_{last}$ is a lower bound for
$t_{BH}$.  If $t_{BH}$ were larger than $t_{last}$ by $\delta t$, then
the number of events expected after $t_{last}$ would be $\delta N
\simeq R(t_{last}) \delta t$.  If Poisson fluctuations caused that
number $\delta N$ to fluctuate to 0, then $t_{last}$ would be smaller
than $t_{BH}$ by about $\delta t$.  This
can only occur for $\delta N \lesssim 1$, or $\delta t \lesssim
1/R(t_{last})$.  Thus, the error in determining the position of a
sharp cutoff is generically of the form $1/R(t_{last})$, i.e., depending
on the number of events $N$ as $1/N$.  For a rate with a tail instead
of a sharp cutoff, the error in determining the offset time scales
instead as $1/\sqrt{N}$; see Ref.~\cite{triang} for a discussion of
the differences.  Furthermore, since $t_{last}$ is always less than
$t_{BH}$, a bias correction $\simeq 1/R(t_{last})$ should be added to
$t_{last}$ to estimate $t_{BH}$.  A more sophisticated treatment of
this problem using order statistics~\cite{Hogg,Eadie} yields the same
scaling results.

Thus, for the Early and Late cases, we find that $t_{BH}$ will be
measured from the charged-current event rate in SK with precision
slightly better than $\simeq 1$ ms and $\simeq 10$ ms, respectively.
These uncertainties on $t_{BH}$ will have a negligible effect on the
mu and tau neutrino mass tests in the lead detector discussed below.


\subsection{Effects of a Nonzero Electron Neutrino Mass}

From the tritium beta decay experiments~\cite{tritium}, the maximum
allowed value of $m_{\nu_e}$ (by CPT, the same as $m_{\bar{\nu}_e}$)
is about 3 eV.  Using Eq.~(\ref{eq:rateafter}), it is straightforward
to calculate the effects of $m_{\nu_e} = 3$ eV on the $\bar{\nu}_e + p
\rightarrow e^+ + n$ event rate after $t_{BH}$ in SK.  Suppose that
$t_{BH}$ were somehow known independently.  By calculating the event
rate and integrating, in the Early case we find 21 events after the
true $t_{BH}$, with delays as large as about 40 ms.  {\it If
unrecognized}, this would bias the extracted $t_{BH}$ to be too large,
and would seriously degrade the mu and tau neutrino mass test (looking
ahead to Fig.~\ref{fig:rateE}).  In the Late case, on the other hand,
we would have only 2.1 events after the true $t_{BH}$, with delays as
large as about 20 ms, with less effect on the mu and tau neutrino mass
test (see Fig.~\ref{fig:rateL}).

However, this potential problem in defining $t_{BH}$ due to the
unknown electron neutrino mass can easily be avoided.  In the reaction
$\bar{\nu}_e + p \rightarrow e^+ + n$ in a \v{C}erenkov detector like
SK, it is possible to measure the neutrino energy by measuring the
positron energy and angle.  At these energies,
\begin{equation}
E_\nu \simeq (E_e + 1.3) 
\left[1 + \frac{E_e}{M} (1 - \cos\theta)\right]\,,
\end{equation}
where $E_e$ is the positron total energy in MeV, $M$ is the nucleon
mass, and $\cos\theta$ is for the positron along the neutrino
direction.  This follows from the two-body kinematics and the small
neutron recoil; the full expression is given in Ref.~\cite{invbeta}.
From Eq.~(\ref{eq:delay}), different neutrino energies correspond to
different delays.  At a given time after $t_{BH}$, only energies low
enough to have caused a delay that large are allowed.  The maximum
allowed energy, Eq.~(\ref{eq:emax}), falls very quickly after $t_{BH}$,
as $1/\sqrt{t - t_{BH}}$.  Thus, different ranges of neutrino energy
will be terminated at different times after $t_{BH}$, and these can be
separated {\it experimentally}.

In general, one would use the event rate as a function of time and
energy, Eq.~(\ref{eq:ratebefore}) and Eq.~(\ref{eq:rateafter}), to
make an unbinned maximum-likelihood fit to the measured neutrino
energies and times to simultaneously measure both $m_{\nu_e}$ and
$t_{BH}$.  However, even without doing that, we can still get a good
idea of how well we can measure $m_{\nu_e}$ and $t_{BH}$ by splitting
the $\bar{\nu}_e + p \rightarrow e^+ + n$ data into different ranges
of neutrino energy, which we define as:
\begin{eqnarray}
& {\rm Low:\ \ \ }  &  0 \le E \le 11.3 {\rm\ MeV}, \nonumber \\
& {\rm Mid: \ \ \ } & 11.3 \le E \le 30 {\rm\ MeV}, \nonumber \\
& {\rm High:\ \ \ } & 30 \le E \le \infty {\rm\ MeV}. \nonumber
\end{eqnarray}
The Low group must be excluded from consideration because these events
have positron total energy less than 10 MeV, and can be confused with
the 5 -- 10 MeV gammas from the neutral-current reaction $\nu +
^{16}{\rm O} \rightarrow \nu + \gamma + X$, where $X$ is either $n +
^{15}{\rm O}$ or $p + ^{15}{\rm N}$~\cite{oxygen}.  In that energy
range, one would not be able to distinguish delay effects due to
$m_{\nu_e}$ or $m_{\nu_\mu}$ and $m_{\nu_\tau}$.  Generally speaking,
$m_{\nu_e}$ and $t_{BH}$ are correlated when extracted from the data
(since $m_{\nu_e} > 0$ has the effect of apparently increasing
$t_{BH}$).  However, the High group has much less delay and will thus
primarily be sensitive to $t_{BH}$.  Then the Mid group will
principally be sensitive to $m_{\nu_e}$, by counting events delayed past
the $t_{BH}$ determined by the High group.

\begin{figure}[t]
\centerline{\epsfxsize=3.25in \epsfbox{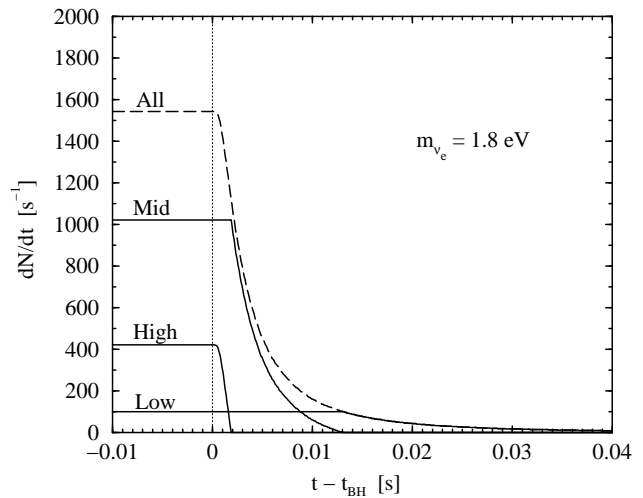}}
\caption{The event rate due to $\bar{\nu}_e + p \rightarrow e^+ + n$
in SK, in the Early case, with an assumed distance of 10 kpc.
Note that only the rate after about $t_{BH}$ is shown, and that the
range of $t - t_{BH}$ is very short.  We took $m_{\nu_e} = 1.8$ eV,
which is close to the minimum mass that can be discerned from this data.
The labels ``Low'' (contains 2.4 events past the true $t_{BH}$),
``Mid'' (4.8 events), ``High'' (0.5 events), and ``All'' (7.7 events)
refer to ranges of neutrino energy defined in the text.}
\label{fig:nuetest}
\end{figure}

In Fig.~\ref{fig:nuetest}, we show such a possible analysis for the
case of $m_{\nu_e} = 1.8$ eV, in the Early case.  The numbers of
events after the true $t_{BH}$ are: 2.4 (Low), 4.8 (Mid), and 0.5
(High).  Since in the High group, the number of events in the tail is
$\lesssim 1$, the cutoff appears sharp and the time of the last event
(after the bias correction) specifies $t_{BH}$ to within the
reciprocal of the event rate at the cutoff, i.e., about 2 ms.  This
uncertainty affects the expected number in the Mid group by at most $\pm 2$
events.  Even in this case, one can still reliably see a few delayed
counts after the measured $t_{BH}$, enough to establish a nonzero mass
(the statistics are discussed in detail in Section V).  We have
ignored the 0.4 events expected during the 0.5 ms duration of the cutoff.

Thus, in the Early case it will be realistically possible to probe
electron neutrino masses as small as about 1.8 eV in SK.  The error on
the time $t_{BH}$ extracted from the same data is not as large as the
possible delays ($\simeq 10$ ms, see Fig.~\ref{fig:nuetest}), but
instead depends on the statistics of the High data.  Though from the
High data alone the error on $t_{BH}$ is about 2 ms, we anticipate
that a more sophisticated fit to all of the data will reduce the error
somewhat, to about 1 ms.  The smallest detectable $m_{\nu_e}$ could
probably also be improved slightly.

In the Late case, the laboratory bound of 3 eV on the electron
neutrino mass will generally be stronger than that derived from the
charged-current signal, and $t_{BH}$ will be measured to about 10 ms.


\section{Measurement of the Mu and Tau Neutrino Masses}


\subsection{General Framework}

The basic signature of a mu or tau neutrino mass is the observation of
neutral-current events after $t_{BH}$.  If many counts delayed past
$t_{BH}$ were observed, then Eq.~(\ref{eq:rateafter}) could be used to
make an unbinned maximum likelihood fit to the mass based on how the
rate fell off with time.  The only measurable quantities for any
delayed counts are their arrival times and their total number, since
it is not possible to measure the neutrino energy in neutral-current
interactions.  This is simply because not enough kinematic variables
are measured (the outgoing neutrino and the recoiling nucleus are not
detected).  In neutrino-electron scattering, measurement of the
electron energy and angle would allow reconstruction of the neutrino
energy in principle; in practice, the kinematic range of the outgoing
electron angle is less than the angular resolution of the detectors.
Thus it is not possible to select ranges of neutrino energy as in the
$m_{\nu_e}$ measurement.  While that could be done crudely by
exploiting the different response functions of different targets (see
Fig.~3 of Ref.~\cite{JHUproc}), it is not necessary if $t_{BH}$ is
measured independently in SK.  The various neutral-current yields can
also be used to estimate the $\nu_\mu$, $\nu_\tau$, $\bar{\nu}_\mu$,
and $\bar{\nu}_\tau$ temperature $T$ (or, more generally, the spectral
shape).

The test proposed in this paper is to simply count the number of
neutral-current events after $t_{BH}$.  There is a very simple
relation between the number of delayed counts and the mass, which we
quote again because of its importance:
\begin{equation}
N_{del} =
\frac{dN}{dt}(t_{BH}) \times 0.515 \left(\frac{m}{E_c}\right)^2 D\,,
\label{eq:ndel2}
\end{equation}
where the event rate is in s$^{-1}$, and the other units are as in
Eq.~(\ref{eq:delay}).  The first important point is that while there
is a spectrum of energies, {\it only one integral over that spectrum
is important}, i.e., the one that determines $E_c$.  If instead we
were making a maximum likelihood fit to a large number of delayed
counts, the precise way the tail was filled out {\it would} depend on
more details of the shape of $f(E) \sigma(E)$.  The second important
point is that after consideration of both the supernova neutrino model
and the detector properties, the {\it only remaining unknown} is the
neutrino mass.  The cutoff time $t_{BH}$ can be measured in SK.  The
number of neutral-current counts $N_{del}$ will be measured between
$t_{BH}$ and some suitable stopping point that depends on the size of
the possible delay effects and the detector background rate.  The
neutral-current event rate at $t_{BH}$ due to mu and tau neutrinos
will be measured with small error since it can be measured over an
adequately long interval before $t_{BH}$.  As noted, the central
energy $E_c$ is well-approximated by the Gamow peak of the falling
spectrum and the rising cross section.  Thus $E_c$ depends on the
temperature $T$; if not assumed from theory, this can be estimated
from the data, as noted above.  We assume that the distance $D$ can be
determined by consideration of the total yield of events or by
astronomical techniques (although a supernova at more than several kpc
will be optically obscured by dust, it will still be visible at other
wavelengths).

For given measured quantities, the best-fit mass is
\begin{equation}
m = E_c \,
\sqrt{\frac{N_{del}}{0.515 D \; \frac{dN}{dt}(t_{BH})}}\,.
\label{eq:m1}
\end{equation}
In the likely case of no delayed events observed, then the best-fit
mass is obviously $m \simeq 0$.  A limit can be placed on the mass by
considering the largest mass $m_{lim}$ that could have faked the
massless case.  At a chosen confidence level, this depends on the
largest number of events that could have fluctuated down to 0 events.
For example, using Poisson statistics, an expectation of 2.3 delayed
counts fluctuates to 0 less than 10\% of the time.  Then $m_{lim}$ is
obtained with Eq.~(\ref{eq:m1}) with $N_{del}$ set equal to 2.3.
If $N_{del} > 0$ is measured,
Table~\ref{tab:poisson} can be used to deduce the allowed range of the
expected number of counts and hence the neutrino mass.  Since the
fractional error on $N_{del}$ due to Poisson statistics is large
($\simeq 1/\sqrt{2.3} \simeq 65\%$), errors on other inputs are
expected to be irrelevant.  If a large number of counts were measured,
the Poisson relative error would be smaller, and the uncertainties on
the inputs would play a more important role.

Dropping all constants of proportionality, we can also write $m_{lim}$
as
\begin{equation}
m_{lim} \sim E_c \,
\sqrt{\frac{\langle E \rangle \, D}{\sigma_{eff} L_{BH} M_D}}\,.
\label{eq:m2}
\end{equation}
While no longer written in terms of the directly measured quantities,
this has the advantage of showing the dependence on the theoretical
inputs more explicitly.  For a supernova that does not have the sharp
cutoff in the rate characteristic of black hole formation, the
model-independent $\langle t \rangle$ analysis~\cite{SKpaper,SNOpaper}
yields an $m_{lim}$ that is {\it independent} of the distance $D$ and
that scales with the detector mass $M_D$ as
$1/M_D^{1/4}$~\cite{SNOpaper}.  The different scaling with $D$, and
the much more favorable scaling with $M_D$, are consequences of the
sharp cutoff in the neutrino flux in the present case.

\begin{table}[t]
\caption{This table shows how a given measured number of events $N$
determines a range for the allowed expected number of events $\mu$,
using Poisson statistics.  For the first line, $\mu = 2.3$ is the
largest expectation that yields $N = 0$ at least 10\% of the time.
For the second line, $\mu = 0.1$ is the smallest expectation that
yields $N = 1$ (or greater) at least 10\% of the time, and $\mu = 3.9$
is the largest expectation that yields $N = 1$ (or smaller) at least
10\% of the time.  Successive lines are similar.  The best-fit $\mu$
is shown in parentheses.  Using Eq.~(\ref{eq:m1}), which relates the
number of events and the neutrino mass $m$, the corresponding allowed
range in $m$ can be determined.  Figs.~\ref{fig:massE} and \ref{fig:massL}
can be used for the same purpose.}
\begin{center}
\begin{tabular}{|l|l|}
measured number & allowed range of the expected number \\ \hline\hline
N = 0 & $0.0 \le \; \mu \; (\simeq 0.0) \; \le 2.3$ \\ \hline
N = 1 & $0.1 \le \; \mu \; (\simeq 1.0) \; \le 3.9$ \\ \hline
N = 2 & $0.5 \le \; \mu \; (\simeq 2.0) \; \le 5.3$ \\ \hline
N = 3 & $1.1 \le \; \mu \; (\simeq 3.0) \; \le 6.7$ \\ \hline
N = 4 & $1.7 \le \; \mu \; (\simeq 4.0) \; \le 8.0$ \\ \hline
N = 5 & $2.4 \le \; \mu \; (\simeq 5.0) \; \le 9.3$ \\
\end{tabular}
\end{center}
\label{tab:poisson}
\end{table}


\subsection{Supernova Neutrino Detection in Lead}

Recently, there has been discussion of building a large supernova
detector based on $^{208}$Pb~\cite{Hargrove,Smith,Boyd,Zach}.  A lead
detector would observe supernova neutrinos by detecting neutrons
produced through both neutral-current and charged-current interactions
of the neutrinos with the lead nuclei.  The neutrons would be produced
primarily by the neutral-current interactions of $\nu_\mu$,
$\nu_\tau$, $\bar{\nu}_\mu$, and $\bar{\nu}_\tau$, because these have
the highest temperature.  The neutrons could be detected in (for
example) a liquid scintillator doped with $\simeq 0.1\%$ gadolinium,
which has a very large neutron-capture cross section, yielding an 8
MeV gamma cascade.  The neutron capture time in such a doped
scintillator is very short, of order 0.030 ms~\cite{PaloVerde}, much
smaller than the typical mass delays.

A novel scheme based on a clear solution of lead perchlorate is also
being explored~\cite{Doe}.  Neutrons would be detected by the 8.6 MeV
gamma cascade from capture on $^{35}$Cl, and electrons would be
detected by their \v{C}erenkov light.

The neutral-current cross sections for neutrinos and antineutrinos on
$^{208}$Pb have been calculated by Hargrove et al.~\cite{Hargrove} and
Fuller et al.~\cite{Fuller}.  The calculations in this paper are based
on the Fuller et al. cross section.\footnote{A very recent calculation
by Kolbe and Langanke~\cite{Kolbe} suggests a lower neutral-current
cross section for neutrinos on $^{208}$Pb, although the differences
with the standard Fuller, Haxton, and McLaughlin~\cite{Fuller}
cross section remain unexplained.}
While the Hargrove et al. and the Fuller et al. results for the
spectrum-averaged cross sections agree within 20\% at $T = 8$ MeV, the
underlying calculations are quite different.  As discussed, the cross
section uncertainties have only a minor effect on the mass test if
$N_{del}$ is small.  Nevertheless, a laboratory measurement of the
neutrino cross sections on lead (perhaps with the ORLAND
detector~\cite{ORLAND} at the Spallation Neutron Source) would be
valuable.

Hargrove et al. consider only the allowed contribution.  The cross
section is assumed to be dominated by a narrow M1 resonance at 8 MeV,
so that
\begin{equation}
\sigma(E) \sim (E - 8 {\rm\ MeV})^2\,,
\end{equation}
for neutrino energies $E > 8$ MeV.  However, Fuller et al. find that
the cross section is dominated by the first-forbidden contribution
(they also point out some apparent errors in the Hargrove et
al. calculation of the allowed contribution).  Fuller et al. do not
provide the cross sections as a function of neutrino energy, but
instead only provide thermally-averaged results for various assumed
spectra.  However, it is straightforward to make a reasonable fit to
$\sigma(E)$ itself.  The neutral-current cross section is dominated by
excitations to the giant dipole resonance at $80 {\rm\ MeV}/A^{1/3}
\simeq 14 {\rm\ MeV}$.  This is just below the 2-neutron emission
threshold, and they find the 2-neutron emission probability to be very
low ($\lesssim 5\%$ of all neutrons).  The cross section can be fit by
the form
\begin{equation}
\sigma(E) \sim (E - 14 {\rm\ MeV})^2\,,
\end{equation}
for neutrino energies $E > 14$ MeV.  A fit was made to the Fuller et
al. results, summing the allowed and forbidden (for $T = 8$ MeV, the
latter is about 80\% of the total) contributions, and {\it summing}
the results for $\nu$ and $\bar{\nu}$
(for either $\nu_\mu$ or $\nu_\tau$ channel).  
Using this form, the leading
constant was found to be $2.7 \times 10^{-42} {\rm\ cm}^2$.  After
fitting, the thermally-averaged cross sections in the first six
columns of Table~I of Ref.~\cite{Fuller} were matched to better than
10\%.  The 1-neutron spallation probability is approximately
independent of energy over the relevant range, and can be taken to be
0.90.  It should be emphasized that our fits to the cross section and
branching ratio will only be valid over the limited range of energy
that we consider.  For a temperature $T = 8$ MeV, the
thermally-averaged cross section in Eq.~(\ref{eq:sigmaeff}) for the
{\it sum} of $\nu$ and $\bar{\nu}$ 
(again, for either $\nu_\mu$ or $\nu_\tau$)
on $^{208}$Pb, including the
1-neutron spallation probability, is about $760. \times 10^{-42}$
cm$^{2}$.  For a supernova at 10 kpc in which the neutrino fluxes are
not terminated by black hole formation, the number of 1-neutron
neutral-current events due to $\nu_\mu$, $\nu_\tau$, $\bar{\nu}_\mu$,
and $\bar{\nu}_\tau$, all at $T = 8$ MeV, is 455 events in 1 kton of
$^{208}$Pb with perfect neutron detection efficiency, in agreement
with Ref.~\cite{Fuller} (who use $T = 7.9$ MeV).

It should be noted that the calculations above were specifically for
$^{208}$Pb, which is 52\% of the abundance of natural lead.  On
the basis of the Thomas-Reiche-Kuhn sum rule, Fuller et
al.~\cite{Fuller} argue that the total neutral-current neutrino cross
sections at these energies should scale as $\sigma \sim A$, where $A$
is the mass number.  Thus, the total cross sections for the three
isotopes of lead should be very similar.  The position of the giant
dipole resonance changes only as $\sim 1/A^{1/3}$, and the 2-neutron
emission thresholds are 0.7 MeV higher in $^{206}$Pb and $^{207}$Pb;
therefore, 1-neutron emission will also dominate in these isotopes.


\subsection{Results for a Lead Detector}

In this section, we calculate results for a $^{208}$Pb detector that
is specified by the number of events expected for a supernova at 10
kpc in which the neutrino fluxes are not cut off by black hole
formation.  We assume that the detector will have $\simeq 1000$
1-neutron neutral-current events due to $\nu_\mu$, $\nu_\tau$,
$\bar{\nu}_\mu$, and $\bar{\nu}_\tau$ in this case.  A possible design
for a 4-kton lead detector with about this many events is described by
Boyd~\cite{Boyd}.  This design also includes 10 kton of iron, with a
smaller number of neutral-current events (not included in our
calculations).  Further refinements in the cross section and detector
design~\cite{Boyd,Zach} (and hence the neutron detection efficiency)
may affect the mass of lead required to meet the design goal of
$\simeq 1000$ neutral-current events of this type.  Using the Fuller
et al.~\cite{Fuller} cross section, this goal could be met with a 2.2
kton lead detector with perfect neutron detection efficiency.  We
refer to this lead detector, whatever its eventual precise
specifications, as the OMNIS (Observatory for Multiflavor NeutrInos 
from Supernovae) detector.

\begin{figure}[t]
\centerline{\epsfxsize=3.25in \epsfbox{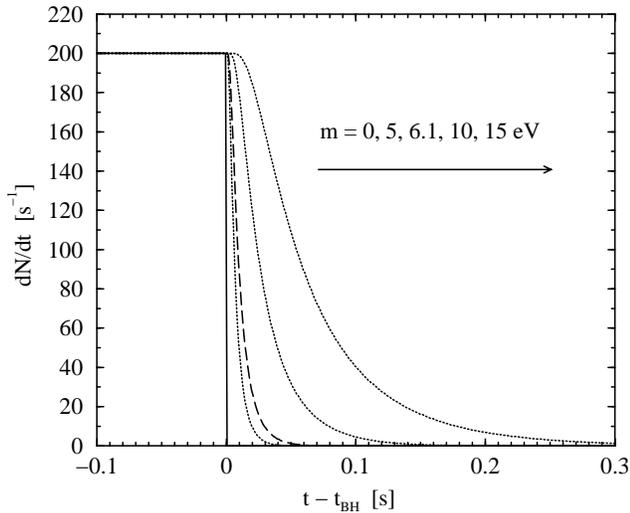}}
\caption{The results for the combined 1-n neutral-current event rate due
to $\nu_\mu$, $\nu_\tau$, $\bar{\nu}_\mu$, and $\bar{\nu}_\tau$ in
OMNIS.  Note that only the rate after about $t_{BH}$ is shown.
The Early case is assumed, with $t_{BH}$ occurring a few ($\sim 1$)
seconds after core collapse, and luminosities of $10^{52}$ erg/s per
flavor at $t_{BH}$.  The assumed distance is 10 kpc.  Before $t_{BH}$,
there are other reactions that produce neutrons; they are not included
here, and those events will have to be statistically subtracted from
the measured neutron rate.  Maximal $\nu_\mu \leftrightarrow \nu_\tau$
mixing with small $\delta m^2$ is assumed, so $m \simeq m_{\nu_2}
\simeq m_{\nu_3}$.  The $m = 0$ case is drawn with a solid line.  The
$m = 6.1$ eV case, with 2.3 events expected in the tail, is the first
case that can be reliably distinguishable from $m = 0$, and is drawn
with a long-dashed line.  The results for other masses are
drawn with dotted lines.}
\label{fig:rateE}
\end{figure}

\begin{figure}[t]
\centerline{\epsfxsize=3.25in \epsfbox{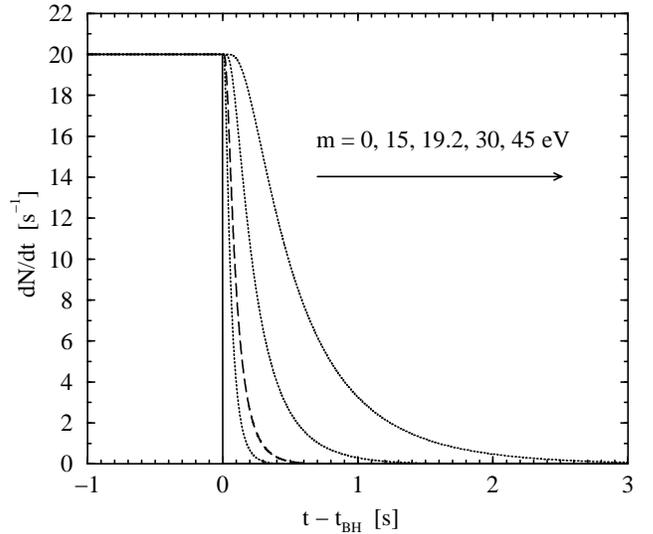}}
\caption{As in Fig.~\ref{fig:rateE}, except that the Late case is
assumed, with $t_{BH}$ occurring within several ($\simeq 10$) seconds
after core collapse, and luminosities of $10^{51}$ erg/s per flavor at
$t_{BH}$.  The $m = 19.2$ eV case, with 2.3 events expected in the
tail, is the first case that can be reliably distinguishable from $m =
0$, and is drawn with a long-dashed line.  Note the changes of scale
on the axes.}
\label{fig:rateL}
\end{figure}

In the following, we assume a supernova distance of 10 kpc.  Using the
product of the thermally-averaged cross section and the branching
ratio given above, the event rates due to neutral-current detection of
$\nu_\mu$, $\nu_\tau$, $\bar{\nu}_\mu$, and $\bar{\nu}_\tau$ can
easily be calculated with Eq.~(\ref{eq:ratebefore}) and
Eq.~(\ref{eq:rateafter}).  These rates are shown in
Fig.~\ref{fig:rateE} and Fig.~\ref{fig:rateL} for the Early and Late
cases.  Recall that the luminosities and cutoff times chosen are
simply examples; in a real case, the relevant quantities will be
measured, not assumed.  In particular, $t_{BH}$ will be measured using
the $\bar{\nu}_e + p \rightarrow e^+ + n$ events in SK.

\begin{figure}[t]
\centerline{\epsfxsize=3.25in \epsfbox{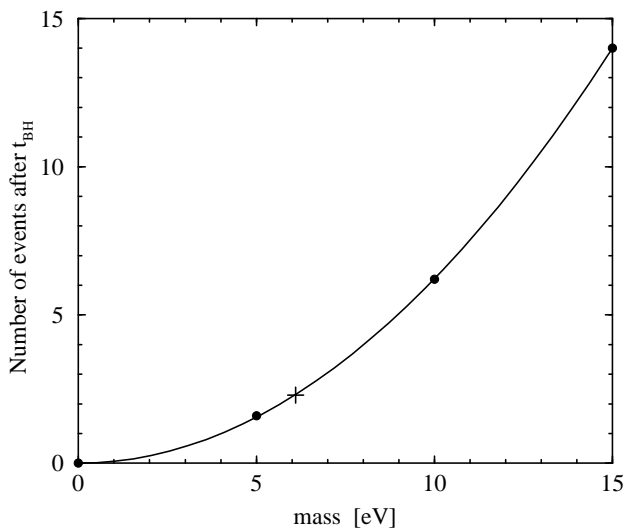}}
\caption{The expected number of delayed counts $N_{del}$ (those after
$t_{BH}$, due to the mass effects) in OMNIS as a function of the
neutrino mass.  The calculation uses the same assumptions as in
Fig.~\ref{fig:rateE}, the Early case.  The points are obtained by
direct numerical integration.  The ``+'' indicates the smallest
discernible mass at the 90\% CL.  The solid line is obtained with
Eq.~(\ref{eq:ndel2}), using $E_c = 40.7$ MeV, the Gamow peak energy.}
\label{fig:massE}
\end{figure}

\begin{figure}[t]
\centerline{\epsfxsize=3.25in \epsfbox{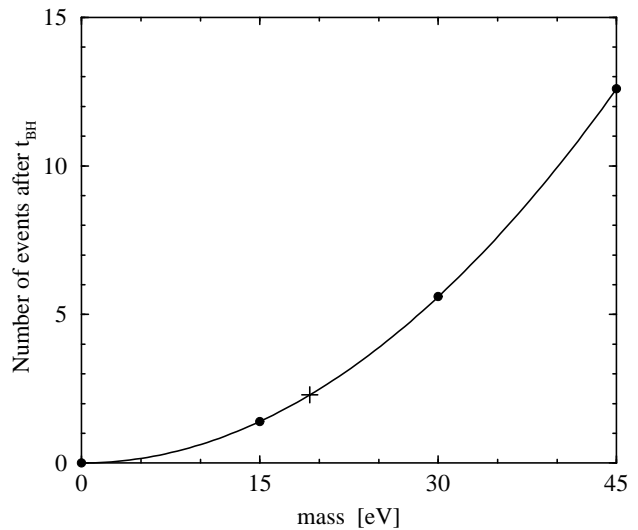}}
\caption{As in Fig.~\ref{fig:massE}, but for the Late case, and with
the assumptions of Fig.~\ref{fig:rateL}.  Note the change in the
horizontal scale.}
\label{fig:massL}
\end{figure}

In Fig.~\ref{fig:massE} for the Early case and in Fig.~\ref{fig:massL}
for the Late case, the number of delayed events $N_{del}$ (that is,
$\nu_\mu$, $\nu_\tau$, $\bar{\nu}_\mu$, and $\bar{\nu}_\tau$ events
after $t_{BH}$) is shown versus the neutrino mass.  The points are from
direct numerical integration of Eq.~(\ref{eq:rateafter}), and the
solid line is the simple analytic result of Eq.~(\ref{eq:ndel2}).
Note that $E_c = 40.7$ MeV is calculated using the Gamow peak of $f(E)
\sigma(E)$, and is not fitted.

In order to use Eq.~(\ref{eq:m1}), a minor correction to the measured
event rate before $t_{BH}$ must be made.  In a lead detector, one
expects to measure just the total neutron rate.  Thus the expected
contributions from the charged-current 1-neutron and 2-neutron events
will have to be statistically subtracted, along with the contributions
of $\nu_e$ and $\bar{\nu}_e$ to the neutral-current rate.  The
subtracted rate of neutrons before $t_{BH}$ is about 20\% of the
total~\cite{Fuller}.
 
The cross section normalization appears only in the event rate, where
it is multiplied by $L_{BH}$, which is a priori unknown.  Only their
product, in the form of the {\it measured} event rate, is needed in
Eq.~(\ref{eq:m1}).  The cross section shape only affects $E_c$.  Using
a Fermi-Dirac spectrum with temperature $T = 8$ MeV, then $E_c \simeq
41$ MeV using the Fuller et al.~\cite{Fuller} cross section given
above, and $E_c \simeq 35$ MeV using the Hargrove et
al.~\cite{Hargrove} cross section given above; this is a negligible
difference.  The spectral temperature $T$ (nominally 8 MeV) of the mu
and tau neutrinos at the time of the cutoff is a priori unknown,
perhaps by $\pm 25\%$, and this also affects $E_c$.  The heavy-flavor
temperature can be estimated from the data by the yields on different
targets (see Fig.~3 of Ref.~\cite{JHUproc}), and this may reduce the
uncertainty on $T$.  Thus, in terms of impact on the measurement of
the neutrino mass, the uncertainties in the thermally-averaged
neutral-current cross section on $^{208}$Pb are of less importance
than the Poisson counting error.

Using Figs.~\ref{fig:massE} and \ref{fig:massL}, 
we obtain mass sensitivity as low as 6.1 eV in
the Early case and 19.2 eV in the Late case.  These are the first
masses that can be reliably discerned (90\% CL) from the massless
case, since they correspond to at least 2.3 expected events after
$t_{BH}$.  Larger masses give even more delayed events, and hence are
easier to measure.  In these results, we have assumed that $\nu_\mu$
and $\nu_\tau$ are maximally mixed, with $\delta m^2 \simeq 10^{-3}$
eV$^2$, as suggested by the atmospheric neutrino results~\cite{SKatm}, so
that both contribute to $N_{del}$.  The results for the neutrino mass
will then apply to the two relevant mass eigenstates.  If we
do not consider this mixing, then perhaps only the tau (or mu) neutrinos
will have a mass and be delayed.  Then $N_{del}$ is half as large as
assumed here, and by Eq.~(\ref{eq:m1}), $m_{lim}$ is $\sqrt{2}$
larger.  Since assuming that only one neutrino is massive is the most
conservative possibility, the deduced limit would in fact apply for
either of the mu and tau neutrino masses.

Finally, we discuss some sources of error for the number of delayed
events $N_{del}$ in a $^{208}$Pb detector, all of which are
negligible.  We ignore possible detector backgrounds over the short
time scale of possible delays.  The duration of the cutoff is about
0.5 ms~\cite{Baumgarte96b}; taking that into account would make
$N_{del}$ larger by $\simeq 0.5 \times 200 \times 0.0005 = 0.05$
events in the Early case and 0.005 events in the Late case.  
As noted, the uncertainty on $t_{BH}$ from SK is assumed to be about 1
ms in the Early case and 10 ms in the Late case.
From Fig.~(\ref{fig:rateE}) and Fig.~(\ref{fig:rateL}), this uncertainty
can be seen to change the expected number $N_{del}$ by $\simeq \pm 200
\times 0.001 = \pm 0.2$ events in the Early case and
$\simeq \pm 20 \times 0.010 = \pm 0.2$ events
in the Late case.  Even with $m_{\nu_e} \lesssim 1.8$ eV determined in
SK, there can still be some $\nu_e$ and $\bar{\nu}_e$ events (charged-
and neutral-current on $^{208}$Pb) after the true $t_{BH}$.  In the
worst case, assuming no tagging on 2-neutron events or events with an
electron, the $\nu_e$ and $\bar{\nu}_e$ events contribute about 20\%
of the total neutron rate before $t_{BH}$.  Assuming $m_{\nu_e} = 1.8$
eV and $E_c \simeq 30$ MeV, then the number of these events after the
true $t_{BH}$ is $\simeq 50 \times 0.515 \times (1.8/30)^2 = 0.09$ in
the Early case and 0.009 in the Late case.  For a larger lead detector
or a closer supernova, some of these errors could become relevant.


\subsection{Results for SNO}

The principal neutral-current reactions available in SNO are $\nu + d
\rightarrow \nu + p + n$ and $\bar{\nu} + d \rightarrow \bar{\nu} + p
+ n$, detected by neutron capture.  For a supernova
at 10 kpc in which the neutrino fluxes are not truncated by black hole
formation, 485 events are expected, of which 400 would be caused by
$\nu_\mu$, $\nu_\tau$, $\bar{\nu}_\mu$, and
$\bar{\nu}_\tau$~\cite{SNOpaper}.  Perfect neutron detection
efficiency is assumed.  Before $t_{BH}$, the neutral-current event
rate due to these flavors may be obtained by scaling the $^{208}$Pb
results by 400/1000, the ratio of the total numbers of events expected
for a supernova that does not form a black hole.  This works simply
because both the event rate before $t_{BH}$ and the total number of events
have the same dependence on $\sigma_{eff}$ and the number of targets.  
Then, using
Eq.~(\ref{eq:m1}) with $E_c = 32$ MeV~\cite{SNOpaper} and $N_{del} =
2.3$, we obtain $m_{lim} = 8$ eV in the Early case and $m_{lim} = 24$
eV in the Late case.

However, it may not be possible to reach this sensitivity in practice
due to the long neutron capture time in heavy water (an exponential
distribution with time constant $\tau_n$).  The value of $\tau_n$
depends on the neutron capture technique: with the dissolved MgCl$_2$
salt, $\tau_n \simeq 4$ ms; with the $^{3}$He counters, $\tau_n \simeq
16$ ms; and with pure D$_2$O, $\tau_n \simeq 35$ ms~\cite{SNO}.  The
effect of this smearing is to delay events after $t_{BH}$ even in the
massless case:
\begin{equation}
N_{del} \rightarrow
N_{del} + \frac{dN}{dt}(t_{BH}) \times \tau_n\,.
\end{equation}
For the Early case, this adds $0.8 (\tau_n/10 {\rm\ ms})$ events after
$t_{BH}$.  Thus, unless the salt is used, the neutrino mass
sensitivity of SNO will be degraded because events after $t_{BH}$ can
be delayed by either $\nu_\mu$ and $\nu_\tau$ mass effects or the
nonzero neutron capture time.


\subsection{Results for SK}

The first set of neutral-current reactions available in SK are those
on $^{16}$O discussed above that yield a 5--10 MeV gamma in the final
state~\cite{oxygen}.  For $\nu_\mu$, $\nu_\tau$, $\bar{\nu}_\mu$, and
$\bar{\nu}_\tau$, 710 events in total are expected for a supernova at
10 kpc~\cite{SKpaper}.  Before $t_{BH}$, the neutral-current event
rate may be obtained by scaling the $^{208}$Pb results by 710/1000.
In practice, this event rate will be obtained from the measured one by
statistically subtracting the comparable rate due to low-energy
$\bar{\nu}_e + p \rightarrow e^+ + n$ events, which are
indistinguishable in SK.  Using Eq.~(\ref{eq:m1}) with $E_c = 60$
MeV~\cite{SKpaper} and $N_{del} = 2.3$, we obtain $m_{lim} = 11$ eV in
the Early case and $m_{lim} = 34$ eV in the Late case.

However, it may not be possible to reach this sensitivity in practice
because of the low-energy $\bar{\nu}_e + p \rightarrow e^+ + n$ events
{\it after} $t_{BH}$, of which there can be as many as 2.4 in the
Early case, due to the limited sensitivity to $m_{\nu_e}$ in SK.
Furthermore, the very steep cross section on $^{16}$O is much more
sensitive to the temperature or the spectral shape in general (see
Fig.~3 of Ref.~\cite{JHUproc}), and so this result is more
model-dependent.  Thus the mu and tau neutrino mass sensitivity of SK
using the neutral-current reactions on $^{16}$O will be limited.

The second set of neutral-current reactions available in SK are $\nu +
e^- \rightarrow \nu + e^-$ and $\bar{\nu} + e^- \rightarrow \bar{\nu}
+ e^-$, for which 120 events due to $\nu_\mu$, $\nu_\tau$,
$\bar{\nu}_\mu$, and $\bar{\nu}_\tau$ are expected for a supernova at
10 kpc~\cite{SKpaper}.  Before $t_{BH}$, the event rate for these
reactions may be obtained by scaling the $^{208}$Pb results by
120/1000.  One must first subtract from the measured event rate events
due to $\nu_e + e^- \rightarrow \nu_e + e^-$, $\bar{\nu}_e + e^-
\rightarrow \bar{\nu}_e + e^-$, and $\bar{\nu}_e + p \rightarrow e^+ +
n$ in the forward cone.  The unwanted events dominate the signal
before $t_{BH}$ by a factor of $\simeq 5$, so the statistical subtraction
will introduce some error.  If this effect can be ignored, then using
Eq.~(\ref{eq:m1}) with $E_c = 25$ MeV~\cite{SKpaper} and $N_{del} =
2.3$, we obtain $m_{lim} = 11$ eV in the Early case and $m_{lim} = 34$
eV in the Late case.

However, it may not be possible to reach this sensitivity in practice,
again because of the limited sensitivity to $m_{\nu_e}$, which can
allow otherwise indistinguishable $\nu_e$ and $\bar{\nu}_e$ events
after $t_{BH}$.  In the Early case, we estimate that there could be
$\simeq 0.9$ such events after $t_{BH}$.  Thus, the mu and tau neutrino
mass sensitivity of SK using the neutral-current reactions on
electrons will also be limited.


\section{Discussion and Conclusions}


\subsection{Distance Dependence of the Neutrino Mass Sensitivity}

\begin{figure}[t]
\centerline{\epsfxsize=3.25in \epsfbox{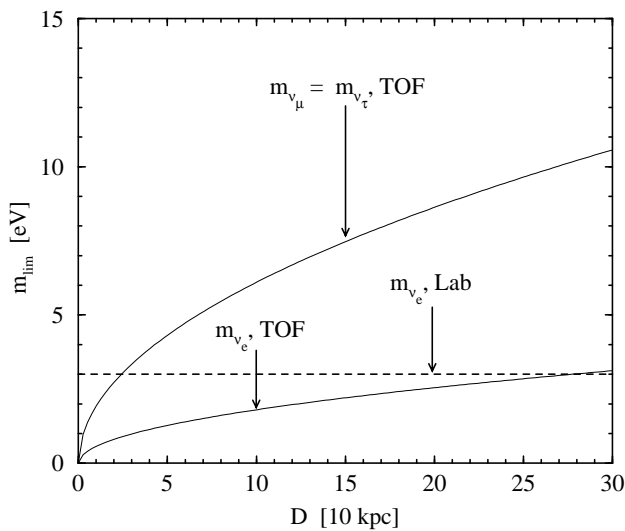}}
\caption{The mass sensitivity as a function of the supernova distance
(solid lines), for the Early case, for $m_{\nu_\mu} \simeq
m_{\nu_\tau}$ measured in the OMNIS detector, and for $m_{\nu_e}$
measured in SK.  This figure is appropriate if $N_{del} = 0$ is
measured and only a limit is being placed on the neutrino mass (if
$N_{del} > 0$ is measured and hence a nonzero mass is discovered, see
Table~I).  The dashed line is the present laboratory upper limit on
$m_{\nu_e}$~\protect\cite{tritium}.  In using Eq.~(\ref{eq:m2}) to
make this figure, we assumed that the events in the $\simeq 0.5$ ms
tail can be disregarded.  Depending on the unknown details of the tail,
this assumption will break down at perhaps $\sim 3$ kpc and the mass
sensitivity will not improve further with decreasing distance.}
\label{fig:ddepE}
\end{figure}

\begin{figure}[t]
\centerline{\epsfxsize=3.25in \epsfbox{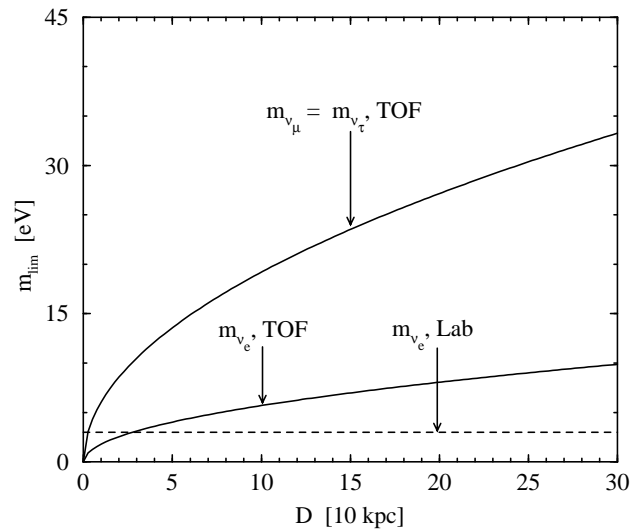}}
\caption{As in Fig.~\ref{fig:ddepE}, but for the Late case.  Because
of the lower luminosity, the mass sensitivity may flatten out only
below about $\sim 1$ kpc.  Note the change in the vertical scale.}
\label{fig:ddepL}
\end{figure}

Throughout this paper, we have assumed that the next Galactic
supernova will be at a distance of 10 kpc.  In the Bahcall-Soneira
Galactic model~\cite{Bahcall1,Bahcall2}, 25\%, 50\%, and 75\% of
supernovae are within about 7, 10, and 14 kpc of Earth, respectively.
If the events during the short ($\simeq 0.5$ ms) cutoff can be
disregarded, then the results for other distances can be scaled with
Eq.~(\ref{eq:m2}), and are shown in Fig.~\ref{fig:ddepE} and
Fig.~\ref{fig:ddepL}.  Other errors, for example the error in
$N_{del}$ that comes from the small error on $t_{BH}$, are independent
of $D$ in their relative importance.

A close supernova at 1 kpc would obviously have 100 times as many
events as we have assumed, and would naively have mass sensitivity
about 3 times better than at 10 kpc, i.e., about 2 eV in the Early
case.  However, there could be a number of events during the short
cutoff that would make defining $t_{BH}$ more difficult than for a
more distant supernova (even assuming that the high event rate in SK
does not saturate the detector).  Assuming a $\simeq 0.5$ ms
duration~\cite{Baumgarte96b}, there could be 40 such events in SK in
the Early case and about 4 in the Late case.  Note that these are
estimated simply by the area of the triangle with height given by the
event rate at $t_{BH}$ and width given by 0.5 ms.  In fact, the
neutrino temperatures are falling rapidly during these 0.5 ms, due to
increasing gravitational redshift; taking that and the detection
threshold into account
would reduce these numbers.  Even if $t_{BH}$ could be defined with
negligible error, there could still be neutral-current events after
$t_{BH}$ due to the $\simeq 0.5$ ms duration of the cutoff: perhaps 5
events in OMNIS in the Early case and 0.5 events in the Late case.
Again, these are conservatively large estimates.
The presence of events during the cutoff would weaken the mass sensitivity,
and it would no longer decrease with decreasing distance.  However,
the real behavior of the luminosity and temperature during the cutoff
is not well known, and further modeling along the lines of Baumgarte
et al.~\cite{Baumgarte96b} is needed.

For an extremely close (and hence rare) supernova, e.g., Betelgeuse at
$\sim 0.1$ kpc, the possibilities are even greater, particularly for
exploring the process of black hole
formation~\cite{Baumgarte96b,Keil}, provided that the neutrino
observatories can accommodate the enormous event rates.

\subsection{Neutrino Oscillations}

While a full discussion of neutrino oscillations is beyond the scope
of this paper, we make a few brief comments.  Oscillations of $\nu_\mu
\leftrightarrow \nu_\tau$ are not important in the sense that these
flavors cannot be distinguished experimentally.  The atmospheric
neutrino results suggest that both are massive, with a small mass
difference and a large mixing angle~\cite{SKatm}; if so, the measured mass
corresponds to the nearly degenerate mass eigenstates.  Oscillations
of $\nu_\mu, \nu_\tau \rightarrow \nu_s$ will decrease the number of
neutral-current events; this is irrelevant in the sense that 
the mass measurement
depends on the measured, not predicted, event rate at $t_{BH}$.
Oscillations of $\nu_\mu,\nu_\tau \leftrightarrow \nu_e$ (and their
antiparticles) can in principle complicate the mass tests.  However,
because of the higher temperature for $\nu_\mu$ and $\nu_\tau$, such
oscillations would greatly increase the number of charged-current
events and would harden the electron or positron spectrum; see, e.g., 
Refs.~\cite{Fuller,SNosc}.  If evidence of such oscillations were seen, the
formalism presented here could easily be enlarged to include
oscillations.  The positron spectrum from $\bar{\nu}_e + p \rightarrow
e^+ + n$ from SN1987A appears to exclude large
$\bar{\nu}_\mu,\bar{\nu}_\tau \leftrightarrow \bar{\nu}_e$
mixing~\cite{87Aosc}.


\subsection{Conclusions}

If a black hole forms early in a core-collapse supernova,
then the fluxes of the various flavors of neutrinos will be
abruptly and simultaneously terminated when the neutrinospheres are
enveloped by the event horizon.  For a massive neutrino, the cutoff in
the arrival time will be delayed by $\Delta t \sim (m/E)^2$ relative
to a massless neutrino.  The SK detector can measure both $t_{BH}$ and
$m_{\nu_e}$ by the arrival times of low- and high-energy $\bar{\nu}_e
+ p \rightarrow e^+ + n$ events, for which the neutrino energies can
be measured.  The mu and tau neutrinos are detectable only by their
neutral-current interactions, in which their energies are not
measured.  However, their masses can be measured
by counting the number of these neutral-current events detected after
$t_{BH}$.

The mass sensitivity depends on the supernova neutrino luminosity
$L_{BH}$ at cutoff, the distance $D$, and the detector used.  For
luminosities of $10^{52}$ erg/s per flavor at cutoff (the Early case),
and a distance of 10 kpc, SK will be able to measure an electron
neutrino mass as small as 1.8 eV and OMNIS would be able to measure
$m_{\nu_\mu} \simeq m_{\nu_\tau}$ as small as about 6 eV.
These results are perhaps even slightly conservative, as the
luminosities in Refs.~\cite{Burrows88,Mezzacappa} were in fact a few
times larger than assumed in the Early case.
As discussed, the mu and tau neutrino masses were assumed to be
degenerate because of the atmospheric neutrino results~\cite{SKatm};
in this case the masses are really those of the relevant mass
eigenstates.

Using the neutral-current channels in SNO and SK, the neutrino mass
sensitivity is nominally $\simeq 10$ eV for each.  However, it appears
that various practical effects will degrade those results.

For other luminosities, distances, and detector masses, the mass
sensitivity scales as in Eq.~(\ref{eq:m2}), i.e.,
\begin{equation}
m_{lim} \sim \sqrt{\frac{D}{L_{BH} M_D}}\,.
\end{equation}
This should be contrasted with the case in which the neutrino
luminosities are not truncated by black hole formation, where $m_{lim}
\sim 1/M_D^{1/4}$ and is independent of $D$.

As we have discussed, there seems to be a good chance that the ongoing
and proposed neutrino detectors can observe the truncation of the
neutrino signals caused by black hole formation in a Galactic
core-collapse supernova.  This would have profound consequences, even
if no delayed events were observed and only limits were placed on the
neutrino masses.
Besides the obvious astrophysical importance of such an observation,
this could improve the limit on the tau
neutrino mass by a factor of almost $10^7$.  Moreover, the technique
discussed in this paper is the {\it only} known possibility for {\it
direct} measurement of the $\nu_\mu$ and $\nu_\tau$ masses (either
Dirac or Majorana) in the crucial eV range suggested by the indirect
neutrino mass tests~\cite{cosmo1,cosmo2,Barger} discussed in the Introduction.


\section*{ACKNOWLEDGMENTS}

J.F.B. was supported by a Sherman Fairchild Fellowship at Caltech for
the initial portion of this work, and by a David N. Schramm Fellowship
at the NASA/Fermilab Astrophysics Center (supported by the DOE and
NASA under NAG5-7092) for the latter portion.  R.N.B. was supported by
NSF grant PHY-9901241.  A.M. was supported at the Oak Ridge National
Laboratory, managed by UT-Battelle, LLC, for the U.S. Dept. of Energy
under contract DE-AC05-00OR22725.  We thank Felix Boehm, Steve Bruenn,
Will Farr, Josh Grindlay, Manoj Kaplinghat, Gail McLaughlin, Alex
Murphy, Yong-Zhong Qian, Robert Shrock, Petr Vogel, and Jerry
Wasserburg for discussions.



\end{document}